\documentclass[10pt,letterpaper]{article}
\usepackage[top=0.85in,left=1.25in,footskip=0.75in]{geometry}

\usepackage{amsmath}
\usepackage{amssymb}
\usepackage{graphicx} 

\newcommand{\R}{\mathbb R}

\usepackage{changepage}

\usepackage[utf8]{inputenc}

\usepackage{textcomp,marvosym}

\usepackage{fixltx2e}

\usepackage{amsmath,amssymb}

\usepackage{cite}

\usepackage{nameref,hyperref}

\usepackage{microtype}
\DisableLigatures[f]{encoding = *, family = * }

\usepackage{rotating}

\raggedright
\makeatletter
\renewcommand{\@biblabel}[1]{\quad#1.}
\makeatother

\date{}

\usepackage{lastpage,fancyhdr,graphicx}
\fancyheadoffset[L]{2.25in}
\fancyfootoffset[L]{2.25in}
\begin{document}
\vspace*{0.35in}

\begin{flushleft}
{\Large
\textbf{The undecided have the key: Interaction-driven opinion dynamics in a three state model}
}
\\
Pablo Balenzuela$^{1\ast,\dagger}$,
Juan Pablo Pinasco$^{2,\dagger}$,
Viktoriya Semeshenko$^{3,4,\dagger}$
\\
\bf{1} Departamento de F\'isica, Facultad de Ciencias Exactas y Naturales, Universidad de Buenos Aires and IFIBA, CONICET.
Buenos Aires, Argentina.
\\
\bf{2} Departamento de Matem\'atica, Facultad de Ciencias Exactas y Naturales, Universidad de Buenos Aires  and  IMAS UBA-CONICET
Buenos Aires, Argentina.
\\
\bf{3} Instituto Interdisciplinario de Econom\'ia Pol\'itica (IIEP), Facultad de Ciencias Econ\'omicas,
Universidad de Buenos Aires, Buenos Aires, Argentina
\\
\bf{4} Consejo Nacional de Investigaciones Científicas y Tecnicas (CONICET)
\\
\vspace{0.2cm}
$\ast$ E-mail: balen@df.uba.ar \\
\vspace{0.2cm}
$\dagger$ All authors contribute equally to this work.
\end{flushleft}

\section*{Abstract}

The effects of interpersonal interactions on individual's agreements result in a social aggregation process which is reflected in the formation of  collective states, as for instance, groups of individuals with a similar opinion about a given issue. This field, which has been a longstanding concern of sociologists and psychologists, has been  extended into an area of experimental social psychology, and even has attracted the attention of physicists and mathematicians.
In this article, we present a novel model of opinion formation in which agents may either have a strict
preference for a choice, or be undecided. The opinion shift emerges during interpersonal communications,
as a consequence of a cumulative process of conviction for one of the two extremes opinions through repeated interactions.
There are two main ingredients which play key roles in determining the steady state: the initial fraction of undecided agents and
the conviction's sensitivity in each interaction.  As a function of these two parameters, the model presents a wide range of possible solutions, as for instance, consensus of each opinion, bi-polarisation or convergence of undecided individuals. We found that a minimum fraction of undecided agents is crucial not only for reaching consensus of a given opinion, but also to determine a dominant opinion in a polarised situation.
In order to  gain a deeper comprehension of the dynamics, we also present the theoretical master equations of the model.




\section*{Introduction}
\label{sec:intro}

When a group of inter-related individuals discuss around a given item, they are feasible to
change their initial opinions in order to get closer to or farther from other subjects in the
group. This interpersonal dynamics leads to different collective behaviours which could be
categorised either by consensus or coexistence of opinions. If furthermore the topic is a
binary statement, as for example  a pro-against issue, the coexistence of opinions turns out
to be a polarised state, where a fraction of the group holds a given opinion and the rest the
opposite one.

These kind of situations lead naturally to questions like:  Which are the mechanisms
underlying the formation of these collective states? or, can we predict the final collective
outcomes when different mechanisms compete among them, as for instance if some
individuals tend to agree and others tend to disagree?

A numerical modelling approach could be a powerful tool  in order to face these kind of
questions and  test how a given interaction mechanism between pair of agents lead to the
formation of collective states. The typical approach to model opinion formation is to
assume that an individual either adopts the opinion of a neighbour or decides on an opinion based
on the average of neighbouring opinions. We take a different conjecture, that the process of
opinion formation is an emergent behaviour of an underlying dynamics. Here, we
develop a novel mathematical model based on the combination of interaction-based
conviction accumulating dynamics and a threshold-driven opinion change.

In sociological research, the concept of threshold  was stated in the seminal papers of T.
Schelling \cite{Schelling78} and M. Granovetter \cite{Granovetter78} in order to understand
the micro-macro link and the aggregation processes. From the psychological perspective,
accumulative-threshold models have been successfully used \cite{Vickers,Ratcliff} in order to
understand binary decision making problems.

The formation of opinion's collective states and the underlying mechanism has largely
studied from sociology and social psychology
\cite{Festinger57,Homans51,Heider67,Akersetal79,Bikhchandanieta92} among others. There
are five main theories to explain group opinion dynamics \cite{Friedkin1999,Eysenck2002}: Social comparison theory \cite{Festinger57,Baronetal96,Sanders77};
Information or Persuasive arguments theory (PAT) \cite{Vinokuretal78,Myers82};
Self-categorization theory \cite{Turneretal87,Oakesetal91,Oakesetal94}; Social decision
theory \cite{Davis73,Kerr92,Kameda2005} and Social influence network theory
\cite{Friedkin1999}.
Predominantly, these theories are focused on interpersonal interactions, and it is useful 
because it draws attention to the emergent communication patterns. 
Essentially, it deepens the understanding of group dynamics, in particular when combined with 
an appropriate mathematical model. 

From a theoretical point of view, much of the existing modelling development about opinion
dynamics has been addressed from a physics-based framework, where the behavioural
mechanism of social influence are derived from analogies with physical systems, in particular
spins \cite{Catellanoetal2009,Latane81}. The variety of existing models assume that
individuals hold binary or continuous opinion values (usually lying between -1 and 1), which
are updated over repeated interactions among neighbouring agents. Different models
assume different rules of opinion adaptation, such as imitation \cite{Akersetal79}, averaging
over individuals with similar opinions \cite{Weisbuchetal2002}, the majority
rule\cite{Galam2005}, or more sophisticated rules \cite{MasFlacheHelbing2010,
Sznajdetal2000}. The classical models that are based on the mechanisms of convergence
(after interacting, individuals get more similar) predict consensus. The models which include
mechanism of negative influence (disliking of dissimilar ones) naturally will give place to
bi-polarisation. However, in last years new models with interesting features appeared, as for
instance, a model of continuous opinion based on persuasive arguments theory
\cite{Masetal2013} where bi-polarisation can been produced without including
negative influence explicitly. Another model presented in \cite{LaRoccaetal2014}, explores the
competition between the two antagonist mechanisms, as persuasion and compromise, among
agents with different degree of agreement about a given issue.

In this work we present an agent-based model for a population of interdependent individuals
who simultaneously participate in an artificial interaction process. The individuals can have
two opposing views, and also a third state of indefiniteness. Through the interactions,
individuals with opinions drive the undecided ones towards one of the definite opinions and
also undecided agents can wear down the conviction of decided ones. This gives place to a
random cumulative process where individuals can eventually change their convictions about
the issue in question \cite{LaRoccaetal2014, Souzaetal2012} and, when this conviction
overpasses a certain threshold, the opinion changes. Thus, the opinion dynamics is an
emerging process depending of the underlying evolution of the conviction of each individual
about a given issue. We are mainly interested in the equilibria reached by this population.
We analyse the convergence properties of the system for wide range of parameter's values
and found that three main collective states can be seen: bi-polarisation, consensus and
convergence of undecided agents. Moreover, the model shows that the initial fraction of
undecided agents could be crucial in the determination of consensus or the dominance of a
given opinion in a polarised situation. We also analyse how the model can achieve the
mentioned collective states in a regime of initially low concentration of undecided individuals
and finally we sketch the exact dynamical equations in order to frame the model within the
non-linear and non-local first order differential equation formulation.

The article is organised as follows: the model and the interaction dynamics are presented in section Analysis.
The results are presented and discussed in section Results. We conclude in section Discussion.

\section*{Analysis}
\label{sec:analysis} In order to study the opinion formation process in groups of interrelated
people we develop a simple agent-based model that includes the main relevant features  for
modelling the opinion dynamics: convergence (in subsequent interactions individuals get to
similar others) and negative influence (disliking of dissimilar others).

We consider a population of $N$ individuals $(i=1, 2, \dots, N)$, each one simultaneously
participating in interpersonal communication process, which can be understood as an
exchange of some kind of information between two or more agents over the issue in
question. The population has the following characteristics: each individual is represented as
an agent $i$, with a numerically valued opinion $O(i)$ which stands for agent's posture on a
given issue at period $t$. The opinion variable $O(i)$ can take three values of attitude
towards an  issue: positive $O(i)= +1$, negative\footnote{It is
important to remark that, the negative opinion does not have a negative connotation about
the issue in question. The negative sign is due to the numerical representation of the
opinion. }  $O(i)= -1$, and neutral $O(i)= 0$.

In addition, the agent $i$ has a conviction about the issue in question, represented by a
variable $C(i)$. This conviction variable $C(i)$ could vary between $C_{max}$ and $-C_{max}$,
and represents being fully aligned with the interaction communication or totally opposed
with it.

These two variables are not independent, i.e., if the conviction variable $C(i)$ is greater than
a given positive threshold, $C(i)>C_T$, the agent has a positive opinion $O(i)= +1$, and if it
is less than a given negative threshold, $C(i)< - C_T$, the agent's opinion is negative $O(i)=
-1$. If the conviction held by the agent does not allow him to decide with any definite
position, then the agent is undecided and $O(i)= 0$ (see Figure \ref{fig:schemaUmbral} (a)).

\subsection*{Dynamics}
\label{subsec:dynamics}
We are interested in the equilibria reached by the system when
agents meet and interact in successive periods. We model this as a process where agents
may increase or decrease their convictions depending on whether the opponent in each
discussion has positive or negative opinion. These social interactions produce cumulative
changes that can eventually lead to the change of opinion: a shift in opinion occurs when
that conviction exceeds a certain threshold.

The interaction dynamics between agents is the following: whenever two agents $i$ and $j$
interact, the agents' conviction values $C(i)$ and  $C(j)$ are modified by an amount bounded
by $k \Delta$, depending on the opinion of both individuals. This parameter, $\Delta$, is a
measure of the sensitivity of the agents to a given interaction.

This social influence mechanism acts as by moving the respective convictions from their
existing positions towards new ones depending of the interacting opponent. This way the
opinion shift is not based on the imitation of opinions of the neighbours (like the typical
imitation behaviours), but is due to  a cumulative process of conviction for one of the two
extremes opinions, after repeated interactions.

The presence of the two social ingredients of the interacting dynamics, convergence and
negative influence, will produce an ``anti-flocking\footnote{In flocking, two distant agents
get attracted, but closer ones repel (the birds want to fly together, but not crashing).} effect'':
Distant individuals repel (caused by mechanism of negative influence) if they differ in
opinions and get attracted by each other if their opinions are similar (mechanism of
convergence and social influence). This last is in line with the fact observed by Wood
\cite{Wood2000} where individuals sharing a common attribute tend to get closer in
opinions.

More formally, at each period $t$, the states of the agents after interaction are updated according to the
following rules, also illustrated in  Figure \ref{fig:schemaUmbral} (b) :
\begin{itemize}
	\item If agents share the same opinions, $O_i(t)=O_j(t)$, and $C_i(t)>C_j(t)$ then they get attracted (agents influence each other so that their convictions become more similar)	
		\begin{eqnarray*}
		C_i(t)=C_i(t)-\Delta, \\
		C_j(t)=C_j(t)+\Delta.
		\end{eqnarray*}
	\item If agents hold different opinions, $O_i(t)=+1$, and $O_j(t)=-1$ then they repel
		\begin{eqnarray*}
		C_i(t)=C_i(t)+\Delta, \\
		C_j(t)=C_j(t)-\Delta.
		\end{eqnarray*}
	If one of the agents has an opinion and another one is undecided, then they get attracted.
	In this case the dynamics is asymmetric simulating the fact that is not the same convincing someone who
	does not have any opinion yet or making someone change his opinion:
	\item If $O_i=-1$ and $O_j=0$ then
		\begin{eqnarray*}
		C_i(t)=C_i(t)+\Delta, \\
		C_j(t)=C_j(t)-k\Delta.
		\end{eqnarray*}
	\item If $O_i=+1$ and $O_j=0$ then
		\begin{eqnarray*}
		C_i(t)=C_i(t)-\Delta, \\
		C_j(t)=C_j(t)+k\Delta.
		\end{eqnarray*}
\end{itemize}

In the way the rules are actually written, some anomalous behaviour can arise for large
values of $\Delta$. For instance, if two agents have very similar convictions, they should be
attracted. But if $\Delta$ is larger than the difference of their convictions, it would happen
that after the interaction they will be more distant in their convictions. In order to avoid these
kind of undesirable behaviours, we set $\Delta_{eff}=\frac{|C(i)-C(j)|}{2}$ if $|C(i)-C(j)|<
\Delta$.

The main parameters of the model are:
\begin{itemize}
  \item [-] $\Delta$ is a sensitivity parameter which measures how much the conviction of an agent changes after
  each interaction. The smaller it is, the more agents who share the same opinion are needed to convince him.
  \item [-] $P_0$ is the initial fraction of undecided agents.
  \item [-] $C_T$ is the threshold beyond which the agent is no longer undecided and adopts an opinion.
  \item [-] $k$ is a variable which simulates an asymmetry dynamics between an undecided agent and the one with
  a formed opinion. Values $k>1$ imply that the conviction's change of the undecided is modified by a factor $k$ with respect to the other one.
\end{itemize}

Given the details of the interaction dynamics we are interested in the following question: Do the interactions
among the agents with different opinions bring the group to consensus o bi-polarisation?

We present the results of simulations in the next sections. All the simulations are done for
systems with $N=1000$ agents. Results are averages over ensembles $N_{ev}=1000$
equivalent configurations, corresponding to different realisations of the random initial
conditions.

\section*{Results}
\label{sec:results}

In this section we present the steady states of the model and discuss its properties as a function of the relevant parameters.
The aggregate behaviour is characterised in terms of the fraction of agents who state an opinion $i$, $S_i$, where $i=\{+,-,0\}$.
Given that $S_i(t)$ is a function of time, we call $P_i \equiv S_i(0)$ the initial fraction of agents who has an opinion $i$, and $T_i\equiv S_i(t_{asint})$ the fraction of agents who have come to a definite position on an issue or have "no opinion" at convergence.

There are two features of the model that, a priory,  are preferred to be fixed: \emph{first},
opinions should be equally likely; \emph{second}, it is assumed to be easier for agents who
are undecided to adopt a particular opinion because they have got persuaded, than for those
who have an opinion to turn into "undecided". The first one is implemented in a way that
once the initial fraction of undecided agents, $P_0$, is chosen, the rest of the agents are
equally distributed with both opinions ($\pm 1$). When we change this condition, we will call
$P_+$ the fraction of agents with $O=+1$ after  the undecided are assigned. The second
one is achieved by setting $k=2$. The conviction of each agent is chosen at random from an
uniform distribution within each opinion.

For better description and interpreting interaction effects is especially useful to take in Figure \ref{fig:schemaUmbral}.

\subsection*{Equilibrium States}

We start with analysing the steady states $T_i\equiv S_i(t_{asint})$ as a function of $P_0$
and $\Delta$.  We vary them  $0.01 \le P_0 \le 0.99$ and $0.01 \le \Delta \le 1.0$ with steps
of $0.01$ \footnote[1]{$P_0=0$ and $P_0=1$ are already the equilibrium absorbing states, in
both cases the interaction will not change the opinions of the agents. If initially all agents
are undecided, they will remain undecided. On the contrary, if agents are equally distributed
between the two extreme opinions, the distribution of opinions will remain unchanged.}.
Results are obtained with asynchronous updating where the procedure is iterated until
convergence.

As a result of simulations, the system converges to one of these four equilibrium states:
\begin{description}
\item [(a)] All undecided agents ($O_i=0$ $\forall i$).
\item [(b)] Positive Consensus ($O_i=+1$ $\forall i$).
\item [(c)] Negative Consensus ($O_i=-1$ $\forall i$).
\item [(d)] Bi-Polarisation (fractions of agents having $O=+1$ and $O=-1$).
\end{description}

The steady state is reached when $T_0=1$ or $T_0=0$, i.e. agents are either all undecided or
all have an opinion. Given the constraint  $S_0(t) + S_+(t) + S_-(t)=1$ ($\forall t$), if $T_0=0$,
the only possible equilibria are either consensus of one of the extreme opinion (b-c) or
bi-polarisation (d).

If we look at the system in terms of the fraction of undecided agents, there is a transition
from  $T_0=1$ to $T_0=0$ (i.e. there are no longer agents who have "no opinion"). This
transition depends on the value of $P_0$ and $\Delta$. This result is due to a nonlinear
nature of the underlying opinion forming dynamics, and is different to what is observed in a
reference paper \cite{Vazquez2004}. The authors examined a three-state generalisation of
the voter model where two states ("rightists" and "leftists") are incompatible and interact
with a third state ("centrists") to impose their consensus. In this case the dynamics can settle
in a polarised state consisting only of leftists and rightists.

\begin{figure}
\includegraphics[width=0.9\textwidth]{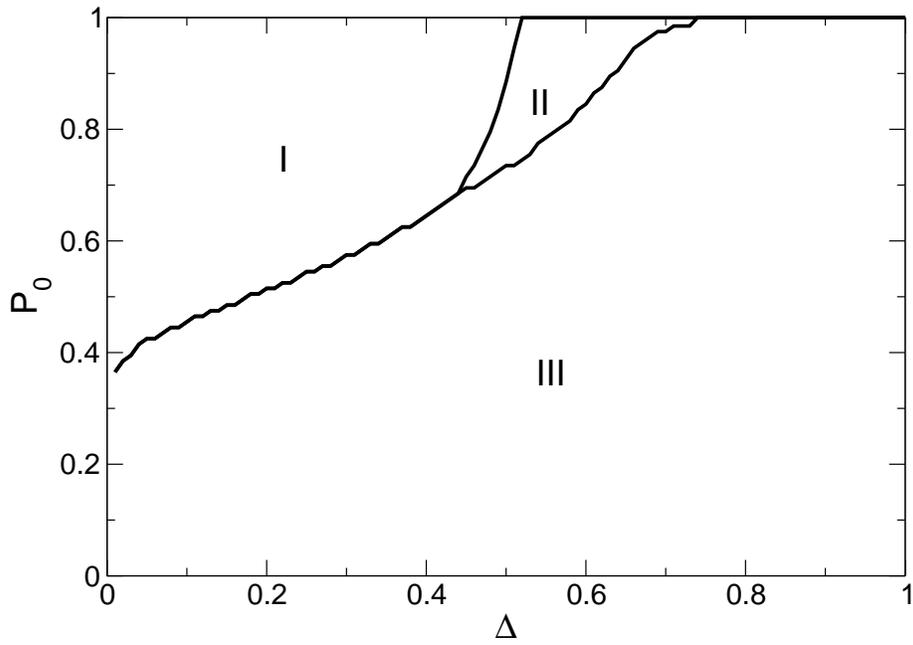}
\caption{Fundamental Phase Diagram: Dominant steady state solution as a function of $P_0$ and  $\Delta$. We identify three different regions: Region I, convergence of undecided; Region II,  consensus of opinions +1/-1 and  Region III,  bi-polarisation. In the simulations $C_T=1$, $k=2$, $C_{max}=3$ were used. }
\label{fig:phasediag}
\end{figure}

Figure \ref{fig:phasediag} shows the Fundamental Phase Diagram ($FPD$) that depicts
existence of different regions of the system under equilibrium. In this Phase Diagram we
analyse the prevalence of each equilibria for every pair of values of $P_0$ and $\Delta$
within the specific range. Be $\langle T_i \rangle$ ($i=\{+,-,0\}$) the ensamble average of
$T_i$, the larger value of $\langle T_i \rangle$ will determine the dominant solution.

The Phase Diagram exhibits three different regions (see Figure \ref{fig:phasediag}):
\begin{description}
\item [Region $I$,] delimited by large values of $P_0$ and small values of $\Delta$, where the equilibrium state is characterised by the convergence of undecided agents (i.e. $O=0$ $\forall i$).
\item [Region $II$,] delimited by large values of $P_0$ and intermediate values of $\Delta$, where the equilibrium state is characterised by the consensus of one of the extreme opinions ($O=+1$ or $O=-1$). This is the most unusual result because it is accompanied with the implicit symmetry breaking in this region.
\item [Region $III$,] delimited by low values of $P_0$ when $\Delta$ is small and by all values of $P_0$ when $\Delta$ is large, where the steady state is defined by a bi-polarisation of opinion, i.e. fractions of agents having $O=+1$ and $O=-1$.
\end{description}

\begin{figure}
\includegraphics[width=0.7\textwidth,angle=270,keepaspectratio,clip]{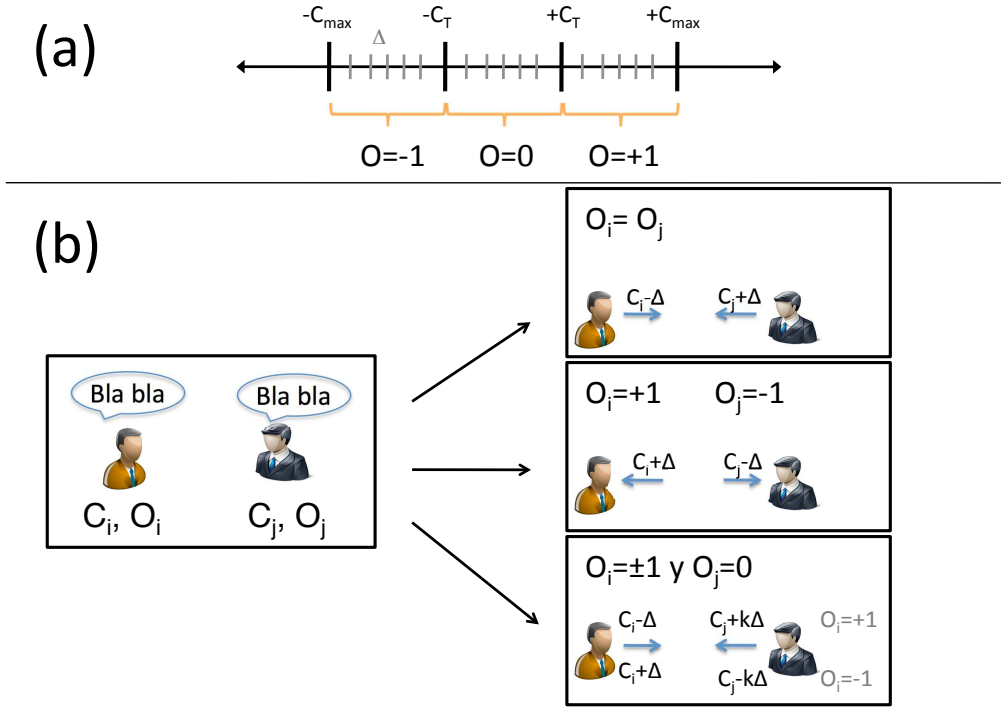}
\caption{Schematic representation of the opinion dynamics. Panel (a): Relation between conviction ($C$) and opinion ($O$). Each conviction interval ($[-C_{max,C-_T}],[-C_T,C_T],[C_T,C_{max}]$) defines one of the three opinion values ($O=-1,0,+1$ respectively). Moreover, each interval is divided in sub-intervals of length $\Delta$. For our simulations we choose $C_T=1$ and $|C_{max}|=3$ (in Figure \ref{fig:var_kthreshold} $C_T$ takes other values also) such as all conviction interval has the same length. Panel (b): Description of how conviction is modified by the pairwise interaction dynamics for three different cases: Same opinion (top panel), opposite opinions (middle panel) and defined opinion vs undecided.}
\label{fig:schemaUmbral}
\end{figure}

Figure \ref{fig:schemaUmbral} can be used again to illustrate a bit more intuitively how
agents interact through this model. When a pair of facing individuals interact, there are two
mechanisms that take place: the first trying to gather together those who share the same
opinion, and the second that makes agents change opinions. For example, if both agents are
undecided they will keep their positions and get closer in their convictions. If one is
undecided and another has an opinion, both will modify their convictions, but the former will
get more similar to the other one, which may eventually make him adopt a position. Due to
this kind of repeated interactions these agents will eventually get closer to the thresholds
$\{-C_T,C_T\}$  (see Figure \ref{fig:schemaUmbral} (a)). On the other hand, if both have
opinions they will not change them, but they will modify their convictions: either approaching
them (if both share the same opinion) or dissociating them (if otherwise). Synthesising, there
are exist two competing opposite forces: first, which drives agents to the center in the
conviction space (turning them to "undecided"), another which drives agents to its extremes,
forcing bi-polarisation of opinions. The final result of mutual interactions depends on the
initial fraction of undecided agents $P_0$, and the sensitivity parameter $\Delta$ (see Figure
\ref{fig:phasediag}).

When $\Delta$ is small, the situation is dominated by agents that change convictions very
little after pairwise interactions. Thus, many of these interpersonal relationships are needed
in order to force a change in their opinions. Here, the fraction $P_0$ is important because it
determines what will be the final state of this dynamics (all undecided ($\langle T_0 \rangle
=1$) or none ($\langle T_0 \rangle =0$)).

Instead, when $\Delta$ increases, the preference for an agent to adopt extreme opinions
(due to the asymmetry given by $k$, see Figure \ref{fig:schemaUmbral}(b)) is more evident
and is reflected in the fact that the more initially undecided agents are needed in order for
the system to reach the final state with ``all undecided". As a consequence, the border of
Region $I$ grows monotonically with $\Delta$ until enclosing with  $P_0=1$. When $\Delta$
is large, the final state is bi-polarisation (Region $III$) independently of the initial density of
undecided agents.

\subsection*{Consensus and Bi-Polarisation}
Given a global picture of the stable solutions, we look in details at the dynamics which
makes the systems reach these states. Furthermore, we restrict the discussion to Regions
$II$ and $III$ where the equilibrium states are consensus and bi-polarisation, respectively,
because Region $I$ does not represent any interest neither from the social nor the individual
point of view \footnote{For example, the statistics on undecided voters indicate that most
individuals have pre-existing beliefs when it comes to politics, and relatively few people
remain undecided late into high-profile elections\cite{Sidoti2008}}.

Region $III$ deserves a closer look. On average, the distribution of $T_+$ and $T_-$ (the
fraction of agents who have an opinion and the fraction of undecided agents at convergence,
respectively) is $50 \% - 50 \%$, and depends on $P_0$ and $\Delta$.

\begin{figure}
\includegraphics[width=0.8\textwidth,keepaspectratio,clip]{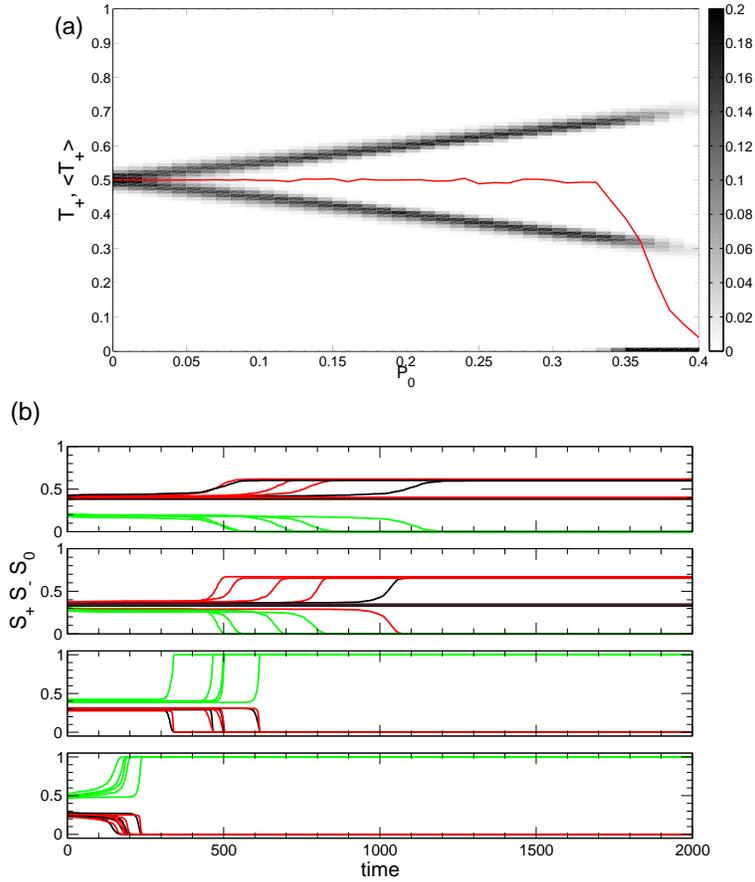}
\includegraphics[width=0.7\textwidth,keepaspectratio,clip]{3B_DynOpPindeVarDi0_01.eps}
\caption{ Bi-polarisation Region. Panel (a): Distribution of positive opinions ($T_+$) and its average ($<T_+>$) as a function of $P_0$ for $\Delta=0.01$. Panel (b): Fraction of agents with opinion $O=+1$ ($S_+(t)$, black), $O=-1$ ($S_-(t)$,red) and undecided ($S_0(t)$, green) as a function of time for different undecided initial concentration (From top to bottom: $P_0$=0.20, $P_0$=0.30, $P_0$=0.40, $P_0$=0.50).}
\label{fig:zona3smasp02}
\end{figure}

On Figure \ref{fig:zona3smasp02} we plot the distribution of positive opinions, $T_+$, and its
average, $\langle T_+ \rangle $ as a function of $P_0$ for different values of $\Delta$. It can
be observed that, as well as the dynamics is confined to Region $III$, the averaged value of
positive opinion is fifty percent, i.e., $\langle T_+ \rangle =0.5$. However, the behaviour of
the distribution of positive (negative) opinion  depends on $P_0$ and $\Delta$. For example,
for $\Delta=0.01$, a clear bimodal distribution emerges (see Figure \ref{fig:zona3smasp02}
(a)). It can be observed that the mean value of each branch depends linearly on $P_0$, either
increasing or decreasing. This behaviour can be understood if we look at Figure
\ref{fig:zona3smasp02}(b), where the dynamics of the fraction of agents with a given opinion
($S_i$, $i=\{+,-,0\}$) is plotted against time for different values of $P_0$. Here it can be seen
that as the system approaches the final state, almost total number of undecided agents
adopts massively either one of the two opinions with the same probability. If for example,
there are initially $20\%$ of undecided agents, it means that there are $40\%$ of agents with
positive opinion and $40\%$ of agents with negative opinion. Due to the dynamics, the
system will converge either to the state with $60\%$ ($T_+=0.60$) or $40\%$ with positive
opinions ($T_+=0.40$). When $P_0=0.1$ the final states for $T_+$ will be $0.45$ or $0.55$,
corresponding to closer values of the mean of the bimodal distribution. But if $P_0=0.30$
then $T_+$ will be $0.35$ or $0.65$ corresponding to the regions where the two branches
are more distant from each other.

\begin{figure}
\includegraphics[width=0.8\textwidth,keepaspectratio,clip]{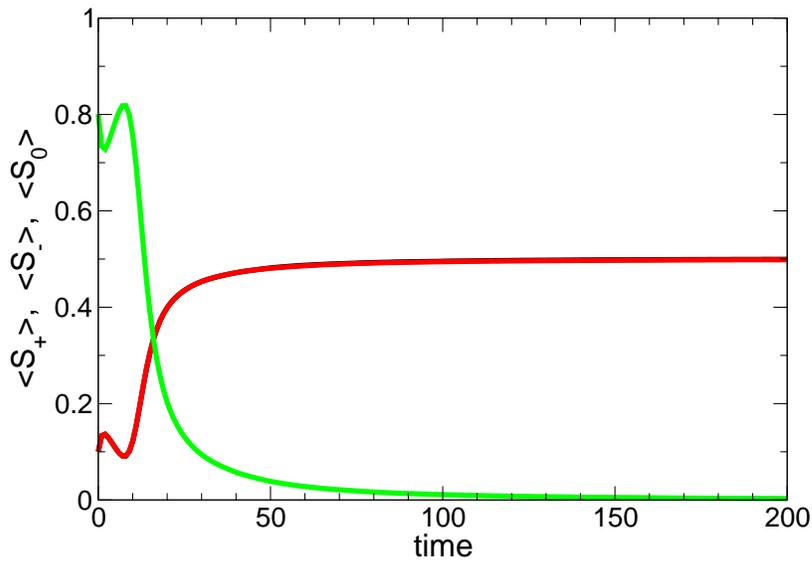}
\caption{Averaged opinion dynamics in region II of the Fundamental Phase Diagram. The plot shows the averaged  evolution in time for each opinion dynamics ($<S_+>$, black, $<S_->$, red and $<S_0>$, green). It can be observed that undecided population grows until  it reach a value above $80\%$ in the first time steps but finally one of the populations with defined opinions becomes dominant with the same probability. Evolutions are averaged over $N_{ev}=10000$. $P_0=0.80$, $\Delta=0.55$. }
\label{fig:zona2dynop}
\end{figure}

Region $II$ is the most interesting one, given that all agents adopt, equally likely, one of the
two extreme opinions. This region expresses the range of parameter values where a given
opinion prevails and it involves a symmetry breaking in the dynamics. In Figure
\ref{fig:zona2dynop} we plot the average opinion dynamics as a function of time for
$P_0=0.80$ and $\Delta=0.55$. If we look at the fraction of undecided agents, it can be seen
that after a slight decrease it reaches more than $80\%$ of the population, and then it
decreases monotonically until disappears. When it happens, just one of the two opinion
survives.

\begin{figure}
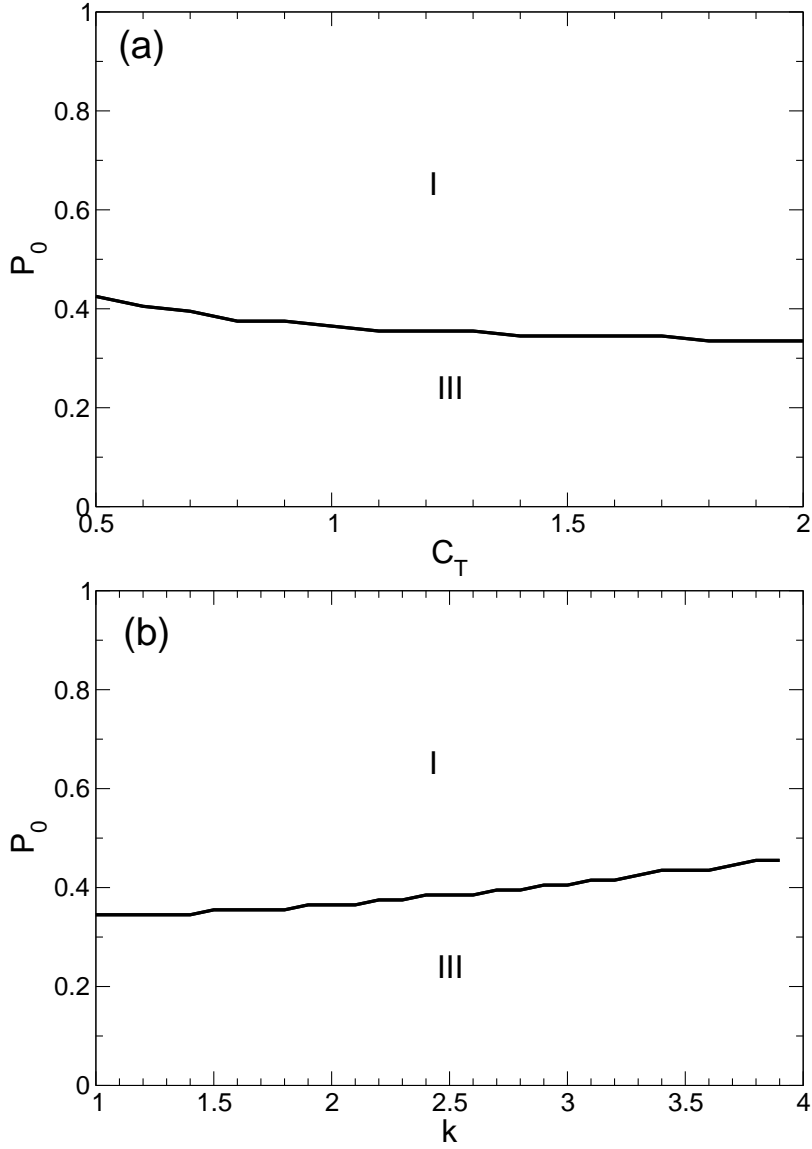

\includegraphics[width=0.8\textwidth,keepaspectratio,clip]{5A_PhaDiagP0_Thrs.eps}
\includegraphics[width=0.8\textwidth,keepaspectratio,clip]{5B_PhaDiagP0_k.eps}
\caption{Alternative Phase Diagram. Panel (a): Regions of dominant solution as a function of  $P_0$ and  $C_T$  for  $\Delta=0.01$. This plot shows that in the low $\Delta$ region and with symmetric distribution of opposite opinions, the systems evolves either to bi-polarisation or convergence of undecided, depending of the initial fraction of undecided individuals, as  have been seen in the Fundamental Phase Diagram (Figure \ref{fig:phasediag}). Panel (b): Same plot  for $P_0$ vs $k$  and  $\Delta=0.01$. In both cases, the main difference of the Fundamental Phase Diagram is the dependence of the transition between Region I and III with $C_T$ and $k$ respectively, as it is explained in the main text.}
\label{fig:var_kthreshold}
\end{figure}

We also explored the impact produced on the Fundamental Phase Diagram when the threshold $C_T$ and
the asymmetry parameter $k$ are modified. Increasing the threshold makes the intermediate
interval of the conviction space, which defines agents to be undecided, to be bigger. Less
undecided are needed initially (at $t=0$) in order to have more of them at the end. This
moves the transition $T_0 =0 \to T_0 =1$ down (see Figure \ref{fig:var_kthreshold} (a)). This in
turn, results in the final disappearance of Region $II$, which produces that Region $I$
becomes dominant in the phase space. Decreasing the threshold produces the opposite
effect: the polarisation Region $III$ becomes bigger.

Instead, with increasing k, grows the tendency to adopt a definite opinion ($+/-1$) after each
interaction, and therefore, more undecided individuals are needed in order to achieve a final
state where they predominate. Thus, this moves the transition up (see Figure
\ref{fig:var_kthreshold} (b)). When the asymmetry parameter is changed, the model moves
between a pragmatic null "symmetric" model ($k=1$) and a more extreme asymmetric case
($k>2$). With large $k$, the polarization region becomes dominant.

\subsection*{How many undecided are relevant?}

In real social situations, the value of $P_0$ depends on the underlying context, and it may be
large or small. Clearly, if we consider examples of voting elections, then assuming very large
values of $P_0$ is less feasible. It is hard to imagine a social system where there is a huge
percentage of undecided voters\footnote{According to the literature and web survey, this
number varies between $10-15 \%$}. In fact, voting must be considered carefully because the
term  ``undecide'' requires correct precision according to Gordon \cite{Gordon2007}, and
Galdi \cite{Galdi2008}.

Instead, social examples of college decision (up to $50\%$ of students enter college as
 ``undecide'' \cite{Gordon2007}) or the choice of major (an estimated $75\%$ of students
change their major at least once before graduation \cite{Gordon2007}) present situations
where large values of $P_0$ are justified.

In the previous section we showed that, for small values of $\Delta$, when the initial fraction of undecided agents, $P_0$
is large, the system presents three solutions, and when $P_0$ is small the equilibrium is
bi-polarisation. Thus, we are interested in the next question: in systems with a relatively
small fraction $P_0$ what should be undertaken in order to obtain the equilibria states
observed previously.

\begin{figure}
\includegraphics[width=0.8\textwidth]{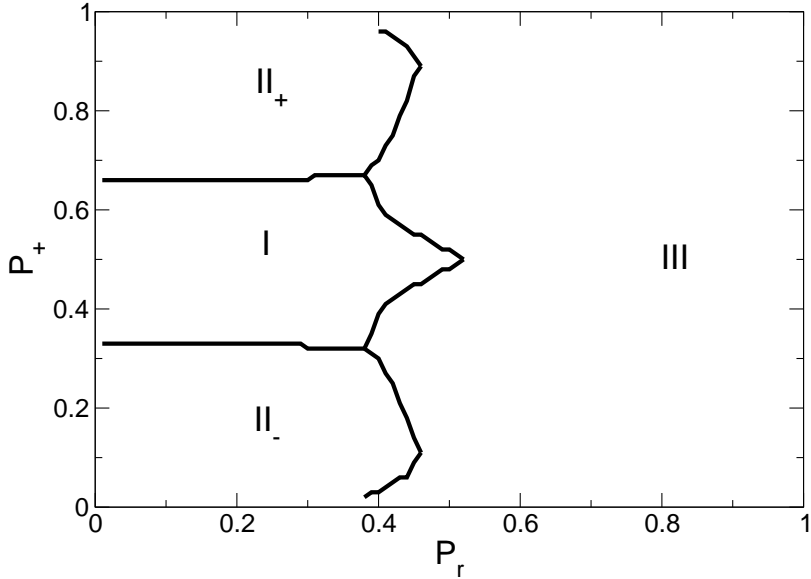}
\caption{Another Alternative Phase Diagram. Regions of dominant solution as a function of  $P_+$ (Bias) and $P_r$ (Repulsion probability) for initially low undecided concentration and small $\Delta$ ($P_0=0.10$ and $\Delta=0.01$). If $P_r=1$, individuals with opposite opinions always repel and the steady state is a polarised situation as we found in the Phase Diagram of Figure \ref{fig:phasediag}. When $P_r < 1$, it can happen that agents with opposite opinions get attracted and the steady state depends on the bias to some opinion. If some of the opinion initially prevails, then the population will go to the consensus of this opinion. Otherwise, a convergence of undecided agents for $P_r < 0.40$ approx. is reached.}
\label{fig:pr_pmas}
\end{figure}

We propose an alternative scenario for the interaction dynamics. Instead of implemented
repulsive effect, allow the individuals with opposite opinions repel with some positive
probability $P_r$, in a similar way it was treated in \cite{LaRoccaetal2014}. The
concentration of undecided is fixed. The Phase Diagram for $P_+$ vs $P_r$ (Repulsion
probability) for initially low concentration of undecided agents and a small value of $\Delta$
($P_0=0.10$ and $\Delta=0.01$) is presented in Figure \ref{fig:pr_pmas}.

When $P_r=1$, agents with opposite opinions always repel and the equilibrium is a polarised
state found in the Phase Diagram on Figure \ref{fig:phasediag}. When $P_r < 1$, agents
sharing opposite opinions may get attracted with  probability $(1-P_r)$, and the steady state
depends on the bias to any opinion. If any of the opinion initially prevails, then the population
will reach the consensus to this opinion. Otherwise, a convergence to ``all undecided" for
$P_r < 0.40$ approx. is reached.

The two scenarios for the interaction dynamics analysed here are interesting from the
social point of view  because, in turn, they correspond to the two different sides of the
public debate: ``Do we learn more from the people with opposite opinions of our own?". On
one side there is a view that we learn much more from people with similar opinions because
we do learn more arguments fortifying that belief and take things as facts. On the other side,
there is a view that we only learn if we look beyond: in a discussion with people with
opposing thoughts we see the different points of view, the exchange of thoughts, etc.

\subsection*{Theoretical Approach}

In order to  gain a deeper comprehension of the dynamics of this model,
we present in this section the master equations corresponding to the dynamics of the
systems. We look at the system as composed by three populations, according the opinions of the
agents. Lets recall than an agent with opinion $O=+1$ has a conviction $C \in
[C_T,C_{max}]$, another with $O=0$ has  $C \in [C_T, -C_T]$ and an individual with $O=-1$
has $C \in [-C_{max}, -C_T]$. We divide the interval corresponding to each opinion in
$M= \Delta^{-1}$ subintervals. Given the evolution of the conviction according to the
interaction-evolution rules detailed in previous sections, the best way to describe the
dynamics of the model is in terms of the density of agents with a given conviction. With this
goal we define this density for $1\le j\le M$ and $t\ge 0$,

$$
 s_+(j,t) = \frac{\#\{ i \in [C_T+(j-1)\Delta, C_T+j\Delta)\}}{N},
$$
 $$
 s_0(j,t) = \frac{\#\{ i \in [-C_T+(j-1)\Delta, -C_T+j\Delta)\}}{N},
$$
 $$
 s_-(j,t) = \frac{\#\{ i \in [-C_T-(j-1)\Delta, -C_T-j\Delta)\}}{N},
$$
which represent the fraction of agent of each opinion with conviction in each interval of
length $\Delta$. In this way we can obtain  a coupled system of $3M$ difference equations
governing  the evolution of the density of agents.

We call $S_i^{k\le j}(t)=\sum_{k=1}^j s_i(j,t)$,  $S_i^{k\ge j}(t)=\sum_{k=j}^M s_i(j,t)$, and
recall that $S_i(t)$ is the fraction of agents with opinion $i$, where $i \in \{+,-,0\}$.

In the following, we omit the variable $t$ in the right hand side of the equations for brevity.
After some characteristic time $\tau$, depending on the rate of the interactions, we have
that the variation on the density of agents is given by the balance between gain and loss
terms. For example, when $O=-1$,  we have
$$
  s_-(j,t+\tau) - s_-(j,t)= G_-(j,t)- L_-(j,t)$$
  where the term $G_-$
corresponds to those agents located at $j-1$ which interact with an agent with opinion
$O=+1$ or an agent with opinion $O=-1$ and a stronger conviction, plus  those agents
located at $j+1$ which interact with an agent with neutral opinion $O=0$ or an agent with
opinion $O=-1$ and a weaker conviction:
$$
G_-(j,t)=2s_-(j-1)[ S_+ + S_-^{k\ge j}]
 + 2s_-(j+1)[ S_0 + S_-^{k\le j}].
$$
On the other hand, the loss term corresponds to interactions between an agent located at $j$
with another agent in any other location,
$$
L_-(j,t)=
   -2s_-(j)[1-s_-(j)].$$

So, in this way we obtain the following system of equations:
$$
\begin{array}{rl}
  s_-(j,t+\tau) - s_-(j,t)= & -2s_-(j)[1-s_-(j)] + 2s_-(j-1)[ S_+ + S_-^{k\ge j}]   \\ &
 + 2s_-(j+1)[ S_0 + S_-^{k\le j}],\\

  s_+(j,t+\tau) - s_+(j,t)= & -2s_+(j)[1-s_+(j)] + 2s_-(j-1)[ S_- + S_+^{k\ge j}]   \\ &
 + 2s_+(j+1)[ S_0 + S_+^{k\le j}],\\

s_0(j,t+\tau) - s_0(j,t)= & -2s_0(j)[1-s_0(j)] + 2s_0(j-1)S_+^{k\ge j} + 2s_0(j+1)S_+^{k\le j}  \\
&
 + 2s_0(j+k)S_- + 2s_0(j-k)S_+ \\

\end{array}$$
for $2<j<M$.

For $j=1,2$ and $M$, the equations are slightly different, as for instance can be seen for
$s_-$:
$$
\begin{array}{rl}
  s_-(1,t+\tau) - s_-(1,t)= & -2s_-(1)[1-s_-(1)] + 2s_0(2) S_-    \\ &
 + 2s_-(2)[ S_0 + S_-^{k\le 1}], \\
  s_-(2,t+\tau) - s_-(2,t)= & -2s_-(2)[1-s_-(2)] + 2s_-(1)[ S_+ + S_-^{k\ge 2}]    \\ &
 + 2s_-(j+1)[ S_0 + S_-^{k\le j}] + 2s_0(1)S_-,\\
  s_-(M,t+\tau) - s_-(M,t)= & -2s_-(M)[1-s_-(M)] + 2s_-(M-1)[ S_+ + s_-(M)] , \\           \\
\end{array}$$

Up to here, we can see that the equations are rather difficult to study, but we can gain more
perspective if we move from this discrete version to a continuous model. We can do that by
introducing (smooth) functions $u_i(x,t)$, $i\in \{0,+,-\}$,  defined for
$(x,t)\in[0,1]\times[0,\infty)$ such that
$$ u_i(j\Delta,t)=s_i(j,t),$$
and we can approximate the spatial partial derivative as
 \begin{equation}\label{forward}
 \Delta \partial_x u_i(j\Delta,t) \approx s_i(j+1,t)-s_i(j,t) \approx    s_i(j,t)-s_i(j-1,t),
 \end{equation}
 and the temporal partial derivative as
$$
 \tau \partial_t   u_i(j\Delta,t) \approx s_i(j,t+\tau)-s_i(j,t) .
 $$

Also,
$$
S_i^{k\ge j}(t) \approx \int_0^{j\Delta} u_i(y,t)dy, \qquad
 S_i^{k\le j}(t) \approx \int_{j\Delta}^1 u_i(y,t)dy,
 $$
We assume that $\tau = \Delta$, which corresponds to a time scaling of the rate of
interactions.

After some algebra, the continuous version of the master equations reads as
\begin{eqnarray*}
\frac12 \partial_t u_-(x,t)
= &\partial_x\Big[   u_-(x,t) \Big( \int_0^x u_-(y,t)dy - \int_x^1 u_-(y,t)dy\Big)  \Big] \\
&
+\partial_x\Big[   u_-(x,t) \Big( \int_0^1 u_0(y,t)dy - \int_0^1 u_+(y,t)dy \Big), \\
\frac12 \partial_t u_+(x,t)
 = &\partial_x\Big[   u_+(x,t) \Big( \int_0^x u_+(y,t)dy - \int_x^1 u_+(y,t)dy\Big)  \Big] \\
&
+\partial_x\Big[  u_+(x,t) \Big( \int_0^1 u_0(y,t)dy - \int_0^1 u_+(y,t)dy \Big),\\
\frac12 \partial_t u_0(x,t)
 = &\partial_x\Big[  u_0(x,t) \Big( \int_0^x u_0(y,t)dy - \int_x^1 u_0(y,t)dy\Big)  \Big] \\
&
+2\partial_x\Big[   u_-(x,t) \Big( \int_0^1 u_0(y,t)dy - \int_0^1 u_+(y,t)dy \Big).
\end{eqnarray*}

The boundary conditions for $s_-$ are given by
\begin{eqnarray*}
u_-(0,t) =  & \Big(\frac{\int_0^1 u_-(y,t)dy}{\int_0^1 u_-(y,t)dy+\int_0^1 u_+(y,t)dy}\Big)  2u_0(0,t),
\\
\partial_x u_-(M,t)= & 0,
\end{eqnarray*}
and the ones corresponding to $u_+$ are similar. The no flux boundary condition at $x=1$
follows from the assumption that the convictions are saturated at $\pm C_{max}$. For $u_0$,
there are two non zero Dirichlet boundary conditions similar to the one for $u_-(0,t)$,
reflecting the incoming agents with opinions $O=\pm 1$.

 In this way we obtain a nonlinear coupled system of  first order differential equations of hyperbolic type including
nonlocal terms and nonlocal boundary conditions.

Let us remark that there are few models of this type, even for a single equation. The works of
Deffuant, Neau, Amblard and Weisbuch, see  \cite{DNAW, WDA} in opinion dynamic models
where only agents with similar opinions can interact, present spatial dependent nonlocal
terms involving a small neighbourhood of a given opinion. However, in these models the
nonlocal terms were replaced by Taylor expansions, and porous media and Fokker-Planck
type equations arise.

True nonlocal terms appear in few works.  We can cite for instance the work of Aletti, Naldi
and Toscani \cite{Aletti}, where the authors studied a model of opinion formation, and the
mean value of the opinions $m(t)$ appears in the transport term,
$$
\partial_tu = \gamma \partial_x\Big( (1-x^2)(x-m(t))u \Big), \qquad (x,t)\in [0,1]\times(0,\infty),
$$
here $\gamma \in [-1,1]$, and $m(t) = \int_0^1 y u(y,t)dy$.
 With a different motivation,
inspired in the dislocation dynamic of crystals,  Ghorbel and Monneau considered in
\cite{Ghorbel} the following equation:
$$
\partial_tu = \Big( c(x)+\int_\R \varphi(u(x-y,t))dy \Big)  \partial_x u
, \qquad (x,t)\in \R\times(0,\infty),
$$
similar equations appeared in continuum mechanics in the theory of deformations and
fractures, see for example \cite{Du} and the references therein.

It is worth noticing that there are few theoretical results and numerical methods for these
problems, which are under active research. These kind of difficulties, as well as the novelty of
the model and the numerical results obtained above, makes that we let the solutions of this
equations for future work that is currently in research. However, few remarks are in order,
which are out of the scope of this paper and deserve a lengthier discussion:

\begin{itemize}
\item We have obtained a  system of transport equations, and the total mass of the solution
    is conserved, that is, for every $t\ge 0$,
    $$u_-(t)+u_+(t)+u_0(t)=1.$$
Some mathematical properties of the solution, like positivity, seems difficult to prove,
although the model clearly generates nonnegative solutions.

\item The partial differential equation  for each opinion has two competing terms: a
    coalescent one, $$
\partial_x\Big[   u_-(x,t) \Big( \int_0^x u_-(y,t)dy - \int_x^1 u_-(y,t)dy\Big)  \Big]
     $$
     depending on the own distribution $u_i$, which  tends to concentrate the agents
    around the mean value of the opinion $i$; and  the other one is a pure transport term,
    $$
    \partial_x\Big[   u_-(x,t) \Big( \int_0^1 u_0(y,t)dy - \int_0^1 u_+(y,t)dy \Big),
$$
    which drives the population to $\pm C_{max}$, $\pm C_T$ depending on the densities of
    the other two populations, as was shown in previous sections.

\item The coalescent terms $\partial_x\Big[   u_i(x,t) \Big( \int_0^x u_i(y,t)dy - \int_x^1
    u_i(y,t)dy\Big)  \Big]$ changes signs, suggesting the existence of shocks (perhaps they
    are smoothed by the transport term). Let us suppose that $\int_0^1 u_i(y,t)dy \sim  C$,
     and let us call $ \mu_i(t)$ the median of the distribution $u_i$, i.e.,
 $$\int_0^{\mu_i(t)} u_i(y,t)dy = \frac{C}{2}.$$
We can rewrite the equation as
    $$ \frac12 \partial_t u_i  =
\partial_x\Big[   u_i(x,t) \Big( 2\int_0^x u_i(y,t)dy
    -C\Big)  \Big],
    $$
    and,  for $x\in (0,\mu(t))$, the characteristic curves travel from left to right, and for
    $x\in(\mu(t), 1)$ they travel from right to left. Hence, we can expect the formation of a
    shock curve along the trajectory of $\mu_i(t)$.
\end{itemize}

\section*{Discussion}
\label{sec:discussion}

In this manuscript we have presented a three-state general opinion formation model based
on the hypothesis of an underlying cumulative conviction-threshold dynamics produced by
repeated interactions in a social environment. Given that each agent can have two
opposite opinions or be in an undecided state, this models apply only for circumstances like a
pro-against issue.

We have presented the main stationary results of the numerical simulations as a function of
the relevant parameters and we found that the model is able to reproduce all the expected
collective behaviours. In their initial formulation the model shows, as a function of $\Delta$,
three different collective states: convergence of undecided, consensus of either of the two
opinions and bi-polarisation. Multiple stable states are only possible for large values of
$P_0$, the initial fraction of undecided agents. A situation where most of individuals are
initially undecided could be presented in low information scenarios about the topic in debate,
as for example the discussion about the environmental impact of fracking in US
\cite{Fracking2014}, or college decisions \cite{Gordon2007}.

A paradigmatic scenario of low values of $P_0$ is, for instance,  a two-candidate political election,
where typical values of undecided in previous poll give percentages around $10-15\%$. The
initial formulation of the model predicts bi-polarisation as the only possible collective state
for this range of values of $P_0$. This is due to the repulsion mechanism assumed in the
model by which two individuals with opposite opinions repel from each other. If we relax this
condition and let the system start from a non-symmetric initial condition as was explained in
previous section, the model can also show the same three mentioned collective states as a
function of $P_r$ and $P_+$, as is shown in Figure \ref{fig:pr_pmas}.

The presented model has very basic assumptions, as for instance, all the agents are identical
(all have the same threshold) and the sensitivity in their convictions after each interaction is
not partner-dependent. Also each agent can interact with everyone and there is no any social
network underlying the interaction among them. But even in this simple scenario, the model
presents a very rich behaviour, with different collective states appearing in different
parameter's region. After the deep analysis presented here, the role and analysis of the
mentioned heterogeneities is let for further work.

Along the results shown in this manuscript, we were able not only to do a detailed analysis of
the numerical simulations of different features of this model, but also to give a glimpse to a
theoretical approach of this model. In the corresponding section, we were able to write down
the exact master equations for the evolution of the probability of having a given conviction
$C$, $s_i$ ($i=+.-,0$) and sketch a set of difference equations for this variables. The
equations are easily put in context in their continuous version. Then, it can be seen that the
master equations are a nonlinear coupled system of  first order differential equations of
hyperbolic type including nonlocal terms and nonlocal boundary conditions. As long as we
know, there are rather few works in solving  this kind of equations and the difficulties  they present to be solved, as was
mentioned in the corresponding section. We let the solution of these equations for future
work that is currently under research.

Finally, we would like to mention the potentiality of this model.  This formulation is general
and covers the social main stream theories of group opinion dynamics. In particular, it is
compatible with the persuasive argument theory. The effect of persuasive arguments can be
modelled by introducing the set of arguments available to individuals for an interpersonal
communication arguments exchange as was done in \cite{Masetal2013}. The model is also
consistent with social decision theory, because from a purely formal point of view, one can
assume any mechanism for opinion revision, be it weighted averages of the group initial
opinions, or imitation dynamics of the neighbours if the networks of interactions is included,
etc. Also, the model may be generalised to self-categorisation theory, similar to Salzarulo
\cite{Salzarulo2006}. We leave these extensions for future work.


\section*{Acknowledgments}

Authors would like to acknowledge financial support of ANPCyT, CONICET and University of Buenos Aires 



%
%
%

\bibliography{OpinionPlos}

\begin{thebibliography}{10}

\bibitem{Akersetal79}
Ronald~L. Akers, Marvin~D. Krohn, Lonn Lanza-Kaduce, and Marcia Radosevich.
\newblock Social learning and deviant behavior: a specific test of a general
  theory.
\newblock {\em American Sociological Review}, 44(4):636--655, 1979.

\bibitem{Aletti}
Giacomo Aletti, Giovanni Naldi, and Giuseppe Toscani.
\newblock First-order continuous models of opinion formation.
\newblock {\em {SIAM} Journal of Applied Mathematics}, 67(3):837--853, 2007.

\bibitem{Baronetal96}
R.~S. Baron, S.~I. Hoppe, C.~F. Kao, B.~Brunsman, B.~Linneweh, and D.~Rogers.
\newblock Social corroboration and opinion extremity.
\newblock {\em Journal of Experimental Social Psychology}, 32:537--60, - 1996.

\bibitem{Bikhchandanieta92}
Sushil Bikhchandani, David Hirshleifer, and Ivo Welch.
\newblock A theory of fads, fashion, custom and cultural change as
  informational cascades.
\newblock {\em Journal of Political Economy}, 100(5):992--1026, 1992.

\bibitem{Fracking2014}
Hilary Boudet, Christopher Clarke, Dylan Bugden, Edward Maibach, Connie
  Roser-Renouf, and Anthony Leiserowitz.
\newblock "fracking" controversy and communication: Using national survey data
  to understand public perceptions of hydraulic fracturing.
\newblock {\em JEPO Energy Policy}, 65:57 -- 67, 2014/// 2014.

\bibitem{Catellanoetal2009}
Claudio Castellano, Santo Fortunato, and Vittorio Loreto.
\newblock Statistical physics of social dynamics.
\newblock {\em Rev. Mod. Phys.}, 81:591--646, May 2009.

\bibitem{Davis73}
James~H. Davis.
\newblock {Group decision and social interaction: A theory of social decision
  schemes}.
\newblock {\em Psychological Review}, 80(2):97--125, March 1973.

\bibitem{DNAW}
Guillaume Deffuant, David Neau, Frederic Amblard, and Gérard Weisbuch.
\newblock Mixing beliefs among interacting agents.
\newblock {\em Adv. Complex Syst.}, 3(1--4):87--98, 2000.

\bibitem{Du}
Qiang Du, James~R. Kamm, Richard~B. Lehoucq, and Michael~L. Parks.
\newblock A new approach for a nonlocal, nonlinear conservation law.
\newblock {\em {SIAM} Journal of Applied Mathematics}, 72(1):464--487, 2012.

\bibitem{Eysenck2002}
Michael~W. Eysenck.
\newblock {\em Simply Psychology}.
\newblock Psychology Press, Madison Avenue, New York, NY, 2002.

\bibitem{Festinger57}
Leon Festinger.
\newblock {\em A theory of cognitive dissonance}.
\newblock Row, Petersen and Company, Evanston, White Plains, 1957.

\bibitem{Friedkin1999}
NE~Friedkin.
\newblock Choice shift and group polarization.
\newblock {\em American Sociological Review}, 64:856--875, 12 1999.

\bibitem{Galam2005}
S~Galam.
\newblock Heterogeneous beliefs, segregation, and extremism in the making of
  public opinions.
\newblock {\em Physical Review E}, 71:046123, 4 2005.

\bibitem{Galdi2008}
Silvia Galdi, Luciano Arcuri, and Bertram Gawronski.
\newblock Automatic mental associations predict future choices of undecided
  decision-makers.
\newblock {\em Science}, 321(5892):1100--1102, 2008.

\bibitem{Ghorbel}
A.~Ghorbel and R{\'{e}}gis Monneau.
\newblock Well-posedness and numerical analysis of a one-dimensional non-local
  transport equation modelling dislocations dynamics.
\newblock {\em Math. Comput.}, 79(271):1535--1564, 2010.

\bibitem{Gordon2007}
V.N. Gordon.
\newblock {\em The Undecided College Student: An Academic and Career Advising
  Challenge}.
\newblock Charles C Thomas Publisher, Limited, 2007.

\bibitem{Granovetter78}
Mark Granovetter.
\newblock Threshold models of collective behavior.
\newblock {\em American Journal of Sociology}, 83(6):1420--1443, 1978.

\bibitem{Kameda2005}
Reid Hastie and Tatsuya Kameda.
\newblock The robust beauty of majority rules in group decisions.
\newblock {\em Psychological Review}, 112(2):494--508, 2005.

\bibitem{Heider67}
F~Heider.
\newblock Attitudes and cognitive organization.
\newblock In M~Fishbein, editor, {\em Readings in attitude theory and
  measurement}, pages 39--41. John Wiley and Sons, Inc, New York, London,
  Sydney, 1967.

\bibitem{Homans51}
George~C Homans.
\newblock {\em The human group}.
\newblock Harcourt Press, New York, 1951.

\bibitem{Kerr92}
Norbert~L Kerr.
\newblock Group decision making at a multialternative task: Extremity,
  interfaction distance, pluralities, and issue importance.
\newblock {\em Organizational Behavior and Human Decision Processes}, 52(1):64
  -- 95, 1992.
\newblock Group Decision Making.

\bibitem{LaRoccaetal2014}
C.~E. La~Rocca, L.~A. Braunstein, and F.~Vazquez.
\newblock The influence of persuasion in opinion formation and polarization.
\newblock {\em EPL (Europhysics Letters)}, 106:40004, 5 2014.

\bibitem{Latane81}
Bibb Latan\'e.
\newblock The psychology of social impact.
\newblock {\em American Psychologist}, 36(4):343--356, 1981.

\bibitem{Masetal2013}
Michael M\"{a}s and Andreas Flache.
\newblock Differentiation without distancing. explaining bi-polarization of
  opinions without negative influence.
\newblock {\em PLoS ONE}, 8:e74516, 11 2013.

\bibitem{MasFlacheHelbing2010}
Michael M\"{a}s, Andreas Flache, and Dirk Helbing.
\newblock Individualization as driving fource of clustering phenomena in
  humans.
\newblock {\em PLoS Comput Biol}, 6:e1000959, 10 2010.

\bibitem{Myers82}
David~G Myers.
\newblock Polarizing effects of social interaction.
\newblock In H~Brandstatter, JH~Davis, and G~Stocker-Kreichgauer, editors, {\em
  Group Decision Making}, pages 125--161. Academic Press, London, 1982.

\bibitem{Oakesetal94}
P.~J. Oakes, S.~A. Haslam, and J.~C. Turner.
\newblock {\em Stereotyping and Social Reality}.
\newblock Oxford, Blackwell, 1994.

\bibitem{Oakesetal91}
P.~J. Oakes, J.~C. Turner, and S.~A. Haslam.
\newblock Perceiving people as group members: The role of fit in the salience
  of social categorizations.
\newblock {\em British Journal of Social Psychology}, 30:125--144, - 1991.

\bibitem{Salzarulo2006}
L~Salzarulo.
\newblock A continuous opinion dynamics model based on the principle of
  meta-contrast.
\newblock {\em Journal of Artificial Societies and Social Simulation}, 9:13, 1
  2006.

\bibitem{Sanders77}
G.~S. Sanders and R.~S. Baron.
\newblock Is social comparison irrelevant for producing choice shifts?
\newblock {\em Journal of Experimental Social Psychology}, 13:303--14, - 1977.

\bibitem{Schelling78}
Thomas~C Schelling.
\newblock {\em Micromotives and Macrobehavior}.
\newblock W. W. Norton Company Ltd, New York, 1978.

\bibitem{Sidoti2008}
L~Sidoti.
\newblock Undecided voters not satisfied with both candidates.
\newblock {\em Associated Press}, 2008.

\bibitem{Ratcliff}
Philip~L Smith and Roger Ratcliff.
\newblock Psychology and neurobiology of simple decisions.
\newblock {\em Trends in Neurosciences}, 27(3):161 -- 168, 2004.

\bibitem{Vickers}
Philip~L Smith and Douglas Vickers.
\newblock The accumulator model of two-choice discrimination.
\newblock {\em Journal of Mathematical Psychology}, 32(2):135--168, 1988.

\bibitem{Souzaetal2012}
S.~R. Souza and S.~Gon\c{c}alves.
\newblock Dynamical model for competing opinions.
\newblock {\em Physical Review E}, 85:056103, 5 2012.

\bibitem{Sznajdetal2000}
K.~Sznajd-Weron and J.~Sznajd.
\newblock Opinion evolution in closed community.
\newblock {\em Int. J. Mod. Phys. C}, 11:1157--1165, 9 2000.

\bibitem{Turneretal87}
J.~C. Turner, M.~A. Hogg, P.J. Oakes, S.~D. Reicher, and M.~S. Wetherell.
\newblock {\em Rediscovering the social group: A self-categorization theory}.
\newblock Oxford, Blackwell, 1987.

\bibitem{Vazquez2004}
F~V\'azquez and S~Redner.
\newblock Ultimate fate of constrained voters.
\newblock {\em Journal of Physics A: Mathematical and General}, 37(35):8479,
  2004.

\bibitem{Vinokuretal78}
Amiram Vinokur and Burnstein Eugene.
\newblock Depolarization of attitudes in groups.
\newblock {\em Journal of Personality and Social Psychology}, 36(8):872--885, 8
  1978.

\bibitem{Weisbuchetal2002}
G.~Weisbuch, G.~Deffuant, F~Amblard, and J-P Nadal.
\newblock Meet, discuss and segregate!
\newblock {\em Complexity}, 7:55--63, Jan-Feb 2002.

\bibitem{WDA}
Gérard Weisbuch, Guillaume Deffuant, and Frédéric Amblard.
\newblock Persuasion dynamics.
\newblock {\em Physica A: Statistical Mechanics and its Applications},
  353(0):555 -- 575, 2005.
\newblock 0378-4371.

\bibitem{Wood2000}
W~Wood.
\newblock Attitude change: persuasion and social influence.
\newblock {\em Annual Review of Psychology}, 51:539--570, 2000.

\end{thebibliography}


\section*{Figure Legends}
%

%
%

\end{document}